\documentclass[10pt, prd, twocolumn
]{revtex4}

\usepackage{amsmath}
\usepackage{amssymb}
\usepackage{graphicx}
\usepackage{bm} 

\begin{document}

\title{Thermodynamics and entropy of self-gravitating matter shells
and black holes in $d$ dimensions}

\author{Rui Andr\'{e}}
\email{rui.andre@tecnico.ulisboa.pt}
\affiliation{Centro de Astrof\'{\i}sica e Gravita\c c\~ao  - CENTRA,
Departamento de F\'{\i}sica, Instituto Superior T\'ecnico - IST,
Universidade de Lisboa - UL, Av. Rovisco Pais 1, 1049-001 Lisboa, Portugal}
\author{Jos\'{e} P. S. Lemos}
\email{joselemos@tecnico.ulisboa.pt}
\affiliation{Centro de Astrof\'{\i}sica e Gravita\c c\~ao  - CENTRA,
Departamento de F\'{\i}sica, Instituto Superior T\'ecnico - IST,
Universidade de Lisboa - UL, Av. Rovisco Pais 1, 1049-001 Lisboa, Portugal}
\author{Gon\c{c}alo M. Quinta}
\email{goncalo.quinta@tecnico.ulisboa.pt}
\affiliation{Centro de Astrof\'{\i}sica e Gravita\c c\~ao  - CENTRA,
Departamento de F\'{\i}sica, Instituto Superior T\'ecnico - IST,
Universidade de Lisboa - UL, Av. Rovisco Pais 1, 1049-001 Lisboa, Portugal}

\begin{abstract}

The thermodynamic properties of self-gravitating spherical thin matter
shells an black holes in $d>4$ dimensions are studied, extending
previous analysis for $d=4$. The shell joins a Minkowski interior to
a Tangherlini exterior, i.e., a Schwarzschild exterior in $d$
dimensions, with $d\geqslant4$, The junction conditions alone together
with the first law of thermodynamics enable one to establish that the
entropy of the thin shell depends only on its own gravitational
radius.  Endowing the shell with a well-defined
power-law temperature
equation of state allows to establish a precise form for the entropy
and to perform a thermodynamic stability analysis for the shell.  A
particularly interesting case is when the shell's temperature has the
Hawking form, i.e., it is inversely proportional to the
shell's gravitational radius.  It is shown in this case that the
shell's heat capacity is positive, and thus the shell is stable, for
shells with radii in-between their own gravitational radius and the
photonic radius, i.e., the radius of circular photon orbits,
reproducing unexpectedly York's thermodynamic stability criterion for
a $d=4$ black hole in the canonical ensemble.  Additionally,
the
Euler equation for the matter
shell is derived,  the Bekenstein and holographic
entropy bounds are studied, and the large $d$ limit
is analyzed. Within this formalism
the thermodynamic properties of black holes can be studied too.
Putting the shell at its own gravitational radius, i.e., in the black
hole situation, obliges one to choose precisely the Hawking
temperature for the shell which in turn yields a black hole
with the
Bekenstein-Hawking entropy.  The stability analysis implies that the
black hole is thermodynamically stable substantiating that in this
configuration our system and York's canonical ensemble black hole are
indeed the same system.  Also relevant is the derivation in a
surprising way of the Smarr formula for black holes in $d$ dimensions.

\end{abstract}

\maketitle



\newpage

\section{Introduction}

Black holes are thermodynamics systems that have an internal energy
\cite{smarr,cbh}, an entropy \cite{bek}, and a temperature
\cite{hawking1}.  A statistical physics thermodynamic treatment can be
given through a path integral approach \cite{hawking2} and
consistently {\bf black holes} can be put in a canonical ensemble by defining a
temperature of a heat bath in a given region of space
\cite{york1,york2,pecalemos}. These works were performed for Schwarzschild and
Reissner-Nordstr\"om black holes in four dimensions.

Self-gravitating matter systems also possess thermodynamic properties.
Perhaps, the simplest self-gravitating matter system is a thin shell.
Thermodynamic studies of thin shells in Schwarzschild and
Reissner-Nordstr\"om four-dimensional spacetimes have been performed
in~\cite{Martinez,LemosQuinta3,Extremal1,Extremal2}
where the entropy and
the stability of the shells were displayed.

Since black holes and self-gravitating matter systems are
thermodynamic systems it is natural to mix both.  This has been done
by putting the combined system of black hole plus matter in a
canonical ensemble \cite{martinesyork}.  One can also conceive of a
black hole surrounded by a thin shell and study the compound system
thermodynamically~\cite{DaviesFordPage,Hiscock}.  One can then
collapse the matter into the initial black hole. The collapse should
be done quasistatically and in thermodynamic equilibrium so that the
whole set up makes sense thermodynamically.  Yet another way is to
suppose no intial black hole and some initial self-gravitating matter
in thermal equilibrium. For instance the thin shells considered in
\cite{Martinez,LemosQuinta3,Extremal1,Extremal2}.  Suppose then the
shell gravitationally collapses again quasistatically up to its own
gravitational radius, i.e., up to the formation of a black hole.  On
the verge of the black hole formation the matter entropy must change
in order to give rise to the final black hole entropy
\cite{LemosQuinta3,Extremal1,Extremal2}.  In this way
one can test how matter entropy transforms into black hole entropy,
see also \cite{pvi}.
For a self-gravitating matter continuum, a generic spacetime
matter structure
that includes thin shells, one can also
address the entropy when the matter is forming a black hole,
a situation that has been
fully developed within the quasiblack hole
formalism \cite{lz1,lz2}. An analogous procedure
to find black hole properties is the
membrane paradigm approach
\cite{pt,lz3,lz4}.

It is surely interesting to see if the thermodynamic properties
for black holes and self-gravitating matter are reproduced in
dimensions different from four and in spacetimes with a cosmological
constant.  In three dimensions, thermodynamic properties of thin
shells in BTZ non-rotating and rotating spacetimes have been
studied~\cite{LemosQuinta2,LemosLopes,Extremal3,Extremal4} with
results that, even in one lower dimension and with the inclusion of a
cosmological constant and rotation, somehow repeat the
four-dimensional results, confirming that the BTZ spacetime is a good
bed test for four-dimensional general relativity.  On the other hand,
the study of higher $d$-dimensional self-gravitating
shells has not been performed.
Since there is the intriguing possibility that the universe has
higher dimensions that might be large or small, in which case they are
hidden at large scales but that pop up at some tiny scales, it is
interesting to study how shells and black holes and their
thermodynamics properties develop in higher $d>4$ dimensions.  Here we
make a thermodynamic study of shells for which the inner spacetime is
spherically symmetric Minkowski and the outer spacetime is a
Tangherlini spacetime, i.e., a Schwarzschild spacetime in
$d$-dimensions, $d\geqslant4$.  We also take the self-gravitating
$d$-dimensional shell to
its own gravitational radius and obtain the thermodynamic properties
of a $d$-dimensional black hole, such as its entropy, its stability,
and the corresponding Smarr formula.

We use known results in $d$
dimensions.  For particle orbits in $d$-dimensions see
\cite{monteiro}, for the Hawking temperature in $d$-dimensions see
\cite{kanti}, and for the Smarr formula in $d$-dimensions see
\cite{smarrd}.  We adopt the thermodynamic formalism presented in
\cite{callen}.  We also study the Bekenstein \cite{Bekenstein} and the
holographic \cite{thooft,Susskind,Hod3} entropy bounds for the
$d$-dimensional shells.
We benefit from the result given in 
\cite{loranz} where it is shown that
to be divergent free quantically black holes must be at
the Hawking temperature.

The paper is organized as follows. 
In Sec.~\ref{m0},  the $d$-dimensional interior  Minkowski
and exterior Schwarzschild, or
Tangherlini, metrics are given, and the
mechanical properties of the self-gravitating
thin shell
that makes the junction of the two
spacetimes are found. 
The thermodynamic properties
of the shell are prescribed, the first law
of thermodynamics applied to the shell
is studied, and a generic expression for the
entropy of the shell is found.
In Sec.~\ref{sec:state},
a power-law equation of state is given
to the temperature, local
thermodynamic stability is analyzed,
the Euler relation is found, 
the holographic and
Bekenstein entropy bounds are
studied, as well as the
 large $d$ case. 
In Sec.~\ref{bh1}, the black 
hole limit is taken and
its properties follow.
In Sec.~\ref{conc1}, conclusions are drawn.

\section{Mechanics and thermodynamics of
self-gravitating static thin shells
in $d$ dimensions}
\label{m0}

\subsection{Mechanics of static thin shells: ADM and rest
masses and
the equation of state for the pressure}
\label{m1}

We write Einstein field equation in $d$ dimensions in the form
\begin{equation}\label{einfeld}
G_{ab} = 8 \pi \,T_{ab}\,,
\end{equation}
where $a,b$ are spacetime indices that run from $0$ to $d-1$, $G_{ab}$
is the Einstein tensor, $T_{ab}$ the energy-momentum tensor, and it is
clear that with this choice for Eq.~(\ref{einfeld}) the
$d$-dimensional Einstein field equation have the same form as the
4-dimensional one. We put the $d$-dimensional gravitational constant
$G_d$ to one and the speed of light $c$ to one.

Consider a spherically symmetric timelike $(d-1)$-hypersurface
$\Sigma$ that partitions a $d$-dimensional spacetime into two
regions. The region on the inside is denoted by an $\rm i$ subscript
sign and the outside region with an $\rm o$ subscript.  The partition
is given by a thin shell and we assume that the inside is a $d$-dimensional
flat
metric with $d\geqslant4$
and the outside is a Tangherlini, or $d$-dimensional
Schwarzschild with $d\geqslant4$, metric.

On the inside the coordinates are ${x_{\rm
i}^\alpha=(t_{\rm i},r,\theta_1,...,\theta_{d-2})}$, where $t_{\rm i}$
is the time coordinate inside, $r$ is the radial coordinate, and
$\theta_k$ are the angular coordinates on a $(d-2)$-dimensional
sphere.  The metric on the flat inside is thus
\begin{equation} \label{LE1}
ds^2_{\rm i} = -F_{\rm i} dt_{\rm i}^2 + \frac{dr^2}{F_{\rm i}} +
r^2 d\Omega^2\,,
\end{equation}
with
\begin{equation} \label{f-}
F_{\rm i} = 1\,,
\end{equation}
and
\begin{equation}\label{solid}
d\Omega^2 = d\theta^2_1 + \sum_{k=2}^{d-2}
\left( \prod_{j=1}^{k-1} \sin^2\theta_j \right)d\theta_k^2\,,
\end{equation}
is the line element on a $(d-2)$-dimensional sphere.

On the
outside the coordinates are 
${x_{\rm o}^\alpha=(t_{\rm o},r,\theta_1,...,\theta_{d-2})}$,
where $t_{\rm o}$ is the time coordinate outside, $r$
is the radial coordinate,
and $\theta_k$ are the angular coordinates.
The metric on the Tangherlini outside is thus 
\begin{equation} \label{LE2}
ds^2_{\rm o} = -F_{\rm o} dt_{\rm o}^2 + \frac{dr^2}{F_{\rm o}} +
r^2 d\Omega^2\,,
\end{equation}
with
\begin{equation} \label{f+}
F_{\rm o} = 1-\frac{{\gamma}m}{r^{d-3}}\,,
\end{equation}
where 
\begin{equation}\label{gamma1}
{\gamma} \equiv \frac{16\pi}{(d-2)\Omega_{d-2}}\,,
\end{equation}
and
\begin{equation}\label{O}
\Omega_{d-2}=\frac{2\pi^{(d-1)/2}}{\Gamma\left(\frac{d-1}{2}
\right)}\,,
\end{equation}
$d\Omega^2$
is the same
line element on a
$(d-2)$-dimensional sphere as in Eq.~(\ref{solid}),
$m$ is the spacetime ADM mass, and $\Gamma$ is the gamma function.
In $d=4$ one has $\Gamma(3/2)=\sqrt\pi/2$,
$\Omega_2=4\pi$ and ${\gamma}=2$.
The spacetime gravitational radius is
\begin{equation}\label{rplus}
r_+=\left( {\gamma}m \right)^{1/(d-3)}\,.
\end{equation}
In $d=4$ one recovers $r_+=2m$.
It is useful to define the gravitational
area $A_+$ given by 
\begin{equation}\label{gar1}
A_+= \Omega_{d-2} r_+^{d-2}\,.
\end{equation}
If the spacetime is a black hole spacetime,
then $r_+$ and $A_+$ are the
horizon radius and the horizon area of the
black hole, respectively.
There is an additional radius that pops out naturally
in our context.
This is the radius of the photon sphere \cite{monteiro}
\begin{equation}\label{rph}
r_{\rm ph}=\left(\frac{d-1}{2}\right)^{\frac1{d-3}}\,.
\end{equation}
For $d=4$ it gives
$3m$, and recall that $r_{\rm ph}=3m$ is the photon sphere,
where photons can have circular trajectories in the
Schwarzschild spacetime. 
The generalization of the photon sphere
radius to
$d$-dimensions is
indeed Eq.~(\ref{rph})
\cite{monteiro}.

The self-gravitating
shell is at the hypersurface $\Sigma$ defined
by
\begin{equation}
r= R\,.
\end{equation}
Letting $\tau$ be
the proper time on the shell,
the shell's
 evolution
is parameterized as
$R(\tau)$, $T_{\rm i}(\tau)\equiv t_{\rm i}|_{\Sigma}$,
and
$T_{\rm o}(\tau)\equiv t_{\rm o}|_{\Sigma}$.
Define the metric and coordinates
on the shell by
$h_{ab}$ and $y^a=(\tau,\theta_1,...,\theta_{d-2})$,
respectively, such that on the shell
the line element 
$ds^2_\Sigma=h_{ab}dy^ady^b$ is
\begin{equation}
ds_{\Sigma}^2 = -d\tau^2 + R^2(\tau) d\Omega^2\,.
\end{equation}
The first junction condition
demands continuity of the metric across
the shell. This is obtained by assuring $[h_{ab}]=0$, where
a square brackets $\left[\;\;\right]$ denotes 
the jump in the quantity across the hypersurface.
The first junction condition then yields 
$- \dot{T_{\rm i}}^2  + \dot R ^2=
-F_{\rm o}\dot{T_{\rm o}}^2
+\frac{\dot R ^2}{F_{\rm o}}=-1$,
where a dot denotes differentiation with respect
to $\tau$ and
we have used $F_{\rm i}=1$.
We can now proceed to the second junction condition.
The shell is assumed to be a perfect fluid
so its stress tensor is given by
$S_{ab} = (\sigma+p) u_a u_b + p
h_{ab}$, where $\sigma$ is the rest energy density,
$p$ is the tangential pressure acting on the $(d-2)$-sphere
at radius $R$, and
$u^a$ is the fluid's $d$-velocity.
Denoting the rest mass of the shell by $M$,
the relation between
$M$, the area $A$ of the shell, and $\sigma$ is 
\begin{equation}
M=\sigma A\,.
\end{equation}
where $A$ is 
\begin{equation}\label{ar1}
A = \Omega_{d-2} R^{d-2}\,.
\end{equation}
The second junction condition is
$S_{ab} = - \frac{1}{8\pi} \left(\left[K_{ab}\right]-
\left[K\right]h_{ab}\right)$,
with $K_{ab}$ and $K$
standing for the extrinsic curvature
and its trace, respectively.
The static case is characterized by 
$\ddot{R}=\dot{R}=0$.
The junction  then yields
\begin{eqnarray}
&m = & M - \frac{{\gamma}M^2}{ 4R^{d-3}}, \label{E1} \\
&p = & \frac{(d-3) {\gamma} M^2}{4(d-2)
\Omega_{d-2} R^{d-2}(R^{d-3} - \frac{{\gamma}M}{2})}\, \label{E2}
\end{eqnarray}
The shell can surely be put at infinity, $R=\infty$. 
On the other hand the
static shell concept only makes sense if the radius
of the shell $R$ bounds from above the  spacetime
gravitational radius $r_+$. For $R= r_+$ the shell turns into a black hole. 
For $R< r_+$ there is no static shell.
Thus, $R$ obeys
\begin{equation}\label{Rr+b}
R\geqslant r_+\,,
\end{equation}
with the inequality being valid up to infinity.
The redshift function $k$
at the shell's position $R$ is a quantity that
appears quite often. It is defined by
\begin{equation}\label{k}
k=\sqrt{ 1- \left( \frac{r_+}{R} \right)^{d-3} }\,.
\end{equation}
We see from Eqs.~(\ref{Rr+b}) and~(\ref{k}) that
\begin{equation}\label{k1}
0\leqslant k\leqslant1\,,
\end{equation}
We can then put the rest mass $M$ and the
tangential pressure
$p$ given in 
Eqs.~(\ref{E1}) and~(\ref{E2}), respectively,  
in terms of the
redshift function $k$ given in Eq.~(\ref{k}) to find
\begin{eqnarray}
&M = \dfrac{2R^{d-3}}{{\gamma}}(1-k), \label{M1} \\
&p = \dfrac{(d-3)(1-k)^2}{16\pi R k}\,. \label{p11}
\end{eqnarray}

\subsection{Thermodynamics on the shell: First law,
functional form of the temperature
equation of state, and entropy}
\label{t1}

Consider the self-gravitating
thin shell to be thermally isolated, i.e.,
it is an adiabatic system. In any
infinitesimal neighborhood of a point in the shell one defines a local
temperature $T$ at the shell, a local entropy density $s$, a local
rest mass density $\sigma$, a local tangential pressure $p$, and a
local element area $a$.  The first law of thermodynamics for this
small region in the shell is $T ds = d\sigma + pda$. This can be
integrated on angles at radius $R$ to give the first law of
thermodynamics for the shell
\begin{equation}\label{1T}
T dS = dM + pdA\,,
\end{equation}
where $S$ is
its entropy, $M$ its rest mass,
$p$ the tangential pressure,
and $A$ its area. We work in the
entropy representation \cite{callen},
i.e., we consider
$S$ as function of $M$ and $A$,
\begin{equation}\label{entrep}
S=S(M,A)\,,
\end{equation}
with $T$ and $p$ being given by equations of
state of the form
$T=T(M,A)$
and $p=p(M,A)$, respectively.
The equation of state for the temperature
$T=T(M,A)$ is free and has to be specified.
The equation of state for the pressure
$p=p(M,A)$ is imposed on us through the
junction conditions and 
is given by Eq.~(\ref{E2}) or Eq.~(\ref{p11}) with the help
of  Eq.~(\ref{ar1}).
Both $T(M,A)$
and $p(M,A)$ are formally given by
$T(M,A)=\left(\frac{\partial S}{\partial M}\right)_{A}$ and 
$p(M,A)=\left(\frac{\partial S}{\partial A}\right)_{M}$.
It is useful to define the inverse temperature
$\beta$, 
\begin{equation}
\label{beta}
\beta=\frac1T\,,
\end{equation}
where also $\beta=\beta(M,A)$.
Equation~(\ref{1T}) is then
$dS = \beta\,dM + \beta\,p\,dA$
and it
can only be an exact differential for the
entropy if the integrability condition
\begin{equation}\label{int1}
\left(\frac{\partial \beta}{\partial A}\right)_{M} =
\left(\frac{\partial (\beta p)}{\partial M}\right)_A
\end{equation}
is satisfied. Then,
given $\beta=\beta(M,A)$ and $p=p(M,A)$
obeying Eq.~(\ref{int1}),
$S=S(M,A)$ in Eq.~(\ref{entrep}) can be determined
explicitly by
integration. 

Using
Eqs.~(\ref{rplus}),~(\ref{ar1}), and~(\ref{E1})
we can make the
thermodynamic variable change ${(M,A)\rightarrow (r_+,R)}$ and
upon using Eq.~(\ref{E2}) or Eq.~(\ref{p11}) find
that the differential for the entropy is given solely by a
differential on the gravitational radius $r_+$,
with Eq.~(\ref{1T})
taking the form
\begin{equation}\label{1T2}
dS =\beta(r_+,R) \frac{d-3}{{\gamma}
 k}r_+^{d-4} dr_+\,.
\end{equation}
In terms of $(r_+, R)$, the integrability condition Eq.~(\ref{int1})
reads
\begin{equation}
\left( \frac{\partial \beta}{\partial R} \right)_{r_+} =
\beta \frac{(d-3)(1-k^2)}{2k^2R},
\end{equation}
which has for solution
an inverse temperature Tolman formula at
the shell's location, i.e.,
\begin{equation}\label{tolman}
\beta(r_+,R) = b(r_+) k(r_+,R),
\end{equation}
where $b(r_+)$ is an arbitrary function
of $r_+$ alone. Since $k\rightarrow 1$ as
$R\rightarrow \infty$, $b$ provides the inverse temperature if the
shell were placed at infinity.
An alternative interpretation
is to consider the Tolman redshift formula.
Suppose that there is some negligible but effective
leaking
in the  form of radiation from the
thermally isolated shell to
infinity. From the Tolman formula
we have that the inverse temperature at a given
radius $r$ of the
leaked radiation is given
by $\beta(r_+,r) = b(r_+) k(r_+,r)$.
At infinity $k(r_+,\infty)=1$ and so
the inverse temperature of the radiation there
is $b(r_+)$. Now, 
inserting 
Eq.~(\ref{tolman})
into the entropy
differential Eq.~(\ref{1T2}), one gets
$
dS = \frac{d-3}{{\gamma}}b(r_+)r_+^{d-4} dr_+
$.
Thus, the total entropy
of the shell
is 
given by
the sum of all the entropy differentials
up to that $r_+$, i.e., 
\begin{equation}\label{S}
S(r_+) = \frac{d-3}{{\gamma}} \int^{r_+}_{0} b(r)\,r^{d-4}dr\,.
\end{equation}
In Eq.~(\ref{S})
the integration constant is fixed under the condition that
$S(0)=S_0$ and we put $S_0=0$.
Eq.~(\ref{S})
provides the equation for the shell's entropy
for any acceptable equation of state for $b(r_+)$
and it shows that 
the entropy
does not depend on 
the shell radius $R$, it 
depends only on
the 
gravitational radius $r_+$. I.e., 
shells with different radius $R$ but
with the same $r_+$ have the same entropy.
This is a known
but nevertheless striking result.

\section{Shells with a power-law equation of state
in $d$-dimensions: Entropy,
local thermodynamic stability, Euler relation, 
entropy bounds, and large $d$}
\label{sec:state}

\subsection{Entropy of a shell with a temperature power-law
equation of state}

We still have the freedom to choose the equation of state
for the inverse temperature of the shell
given in the function $b(r_+)$.
To proceed, we assume as equation of state for $b(r_+)$
a power-law function of the form
\begin{equation}\label{C1}
b(r_+) = 4\pi\eta\frac{a+1}{d-3}\, r_+^{a(d-2)+1},
\end{equation}
where 
$\eta$ and $a$ are free parameters without units and
$ \frac{a+1}{d-3}$ appears for convenience.
We have put the Boltzmann constant equal to one
so that temperature has units of mass.
We also out the
Planck constant $\hbar$ equal to one.
Then the 
Planck length for a
$d$-dimensional spacetime
$l_p=\left(\frac{G_d\hbar}{c^3}\right)^{\frac{1}{d-2}}$ is
one and the 
Planck mass
$m_p=\left(
\frac{c^{5-d}\hbar^{d-3}}{G_d}
\right)^{\frac{1}{d-2}}
$ 
is also one, i.e., all quantities are measured
in Planck units.
The choice in Eq.~(\ref{C1}) for
$b(r_+)$
is analogous to the choice 
in~\cite{Martinez}.
Note that the case $a=0$ is of particular interest
as the inverse temperature has the inverse Hawking temperature
form, it is proportional to $r_+$, see Eq.~(\ref{C1}).
If further we choose $\eta=1$ then
$b(r_+) = \frac{4\pi}{d-3}\, r_+$
and the shell
has precisely the Hawking temperature in $d$-dimensions
\cite{kanti}.

Putting 
Eq.~(\ref{C1})
into Eq.~(\ref{S})
leads to the following expression for the entropy
of the self-gravitating shell
\begin{equation}\label{S1}
S = \frac14\, \eta \, \Omega_{d-2} \,r_+^{(a+1)(d-2)} \,.
\end{equation}
From this expression note that 
\begin{equation}\label{a>-1}
	a>-1\,,
\end{equation}
otherwise the entropy would diverge in the limit $r_+ \rightarrow 0$,
a situation we avoid.
The case $a=0$
that has the Hawking inverse temperature 
form, yields an entropy proportional
to it is proportional to $r_+^{d-2}$, i.e.,
proportional to $A_+$ and so has the
Bekenstein-Hawking form.

\subsection{Intrinsic thermodynamic stability}~\label{IntStability2}

\subsubsection{Generics}

Following~\cite{callen}, one can analyze
thermodynamic local stability of a
system in relation to the entropy fundamental equation $S=S(M,A)$.
Stable solutions are considered usiing Le Chatelier's principle,
which states that 
a stable system will tend to restore its equilibrium
homogeneity state
when a small non-homogeneous change
is performed on it.
The thin
matter shell solution is a good approximation to a layer of
matter with a very small thickness.
Let us divide this layer into an
inner and an outer layer 
with proper mass $M$, say,
each and with no thermic contact. The 
fundamental equation for each layer is $S=S(M,A)$.
So the initial entropy of the total system
is $2S \left( M,A\right)$. Now remove some mass
$\Delta M$ from one subsystem to the other and
get for the entropy of total system
$S(M+\Delta M,A)+S(M-\Delta M,A)$. If the thermic
contact is removed mass flows to one side to the
other and the entropy should increase by
the second law of thermodynamics
to its original value $2S \left( M,A\right)$.
So,
$2S \left( M,A\right)\geqslant
S\left( M + \Delta M, A \right)+S\left(M -
\Delta M , A \right)$.
Taking the limit $\Delta M \rightarrow 0$ it
becomes
\begin{equation}\label{Stab1}
\left( \dfrac{\partial^2 S}{\partial M^2} \right)_A \leqslant 0. 
\end{equation}
The heat capacity is given by
$C_A \equiv \left( \frac{\partial M}{\partial T}
\right)_A = - T^{-2}\left( \frac{\partial^2 S}{\partial M^2}
\right)_A^{-1}$. So, 
Eq.~(\ref{Stab1})
is equivalent to requiring a positive heat capacity at constant
area $C_A$.

Analogously, one can consider
the thermodynamic stability in relation to
the area $A$ and obtain
\begin{equation}\label{Stab2}
\left( \dfrac{\partial^2 S}{\partial A^2} \right)_M \leqslant 0.
\end{equation}

For a small change of both $M$ and $A$ simultaneously, the 
stability condition is
\begin{equation}\label{Stab3}
\left( \frac{\partial^2 S}{\partial M^2} \right) \left( \frac{\partial^2 S}
{\partial A^2} \right) - \left( \frac{\partial^2 S}{\partial M \partial A}
\right)^2 \geqslant 0.
\end{equation}

Note that one can analyze each condition at a time.  Condition
Eq.~(\ref{Stab1}) is the actual stability condition if the self-gravitating
shell is
held at fixed $A$, i.e., at fixed radius $R$.  Condition
Eq.~(\ref{Stab2}) is the actual stability condition if the shell is
held at fixed proper mass $M$.  In the case $A$ and $M$ are free,
condition Eq.~(\ref{Stab3}) also counts and one needs to check it.

\subsubsection{Stability for free proper mass $M$ and at fixed
area $A$,
i.e., at fixed shell radius $R$}
\label{fixedA}

Condition Eq.~(\ref{Stab1}) is the stability condition if the
proper mass $M$ is free to change and the 
shell is held at fixed area $A$, i.e.,  fixed radius $R$.
Since, the heat capacity is given by
$C_A \equiv \left( \frac{\partial M}{\partial T}
\right)_A = -T^{-2}\left( \frac{\partial^2 S}{\partial M^2}
\right)_A^{-1}$, 
Eq.~(\ref{Stab1})
is equivalent to requiring a positive heat capacity at constant
area $C_A$, i.e., 
\begin{equation}\label{Stab11}
C_A\geqslant0\,.
\end{equation}
Equation~(\ref{S1}) together with
Eqs.~(\ref{rplus})
and ~(\ref{E1})
yields
$\left(\frac{\partial^2S}{\partial M^2} \right)_A =
\frac{2(1+a)(d-2)S(M,A)}{(d-3)^2M^2(1-k)^2}
\left[ k^2(d-1+2a(d-2)) -\right. \break \left. 
(d- 3) \right]$.
Thus, Eq.~(\ref{Stab1}), or Eq.~(\ref{Stab11}), gives
\begin{equation}\label{quad1}
k^2(d-1+2a(d-2))-(d-3)\leqslant 0.
\end{equation}
If ${a \leqslant -(d-1)/(2(d-2))}$, Eq.~(\ref{quad1}) is always satisfied.
Since we have imposed $a>-1$, Eq.~(\ref{a>-1}), we have
for $-1<a \leqslant -(d-1)/(2(d-2))$ that Eq.~(\ref{quad1}) is satisfied.
On the other hand, for ${a > -(d-1)/(2(d-2))}$,
Eq.~(\ref{quad1}) is satisfied when
$-k_1 \leqslant k \leqslant k_1$, with
$k_1= \sqrt{\frac{d-3}{d-1 + 2a(d-2) }}$.
Since we have $0\leqslant k$, recall Eq.~(\ref{k}),
this condition can be rewritten as
$0\leqslant k\leqslant k_1$. 
Now, note that 
the expression inside the square root in $k_1$,
i.e., $\frac{d-3}{d-1 + 2a(d-2) }$,
is always greater than one if ${a<-1/(d-2)}$.
Then, since $k\leqslant1$, recall again Eq.~(\ref{k}),
we have that from the equation above,
if ${a\leqslant-1/(d-2)}$,
Eq.~(\ref{quad1}) is always satisfied.
Since we had found
that for $-1<a \leqslant -(d-1)/(2(d-2))$ Eq.~(\ref{quad1})
is satisfied, we can extend this range to 
$-1<a \leqslant-1/(d-2)$.
Now, for 
$a >-1/(d-2)$ the expression inside the square root in $k_1$,
i.e., $\frac{d-3}{d-1 + 2a(d-2) }$,
is always smaller than one. This imposes a requirement
on $k$ indeed, i.e., $k\leqslant k_1$, with $k_1\leqslant1$.{}

In brief, stability for free proper mass $M$
and at constant area $A$ means
that Eq.~(\ref{Stab1}) holds which in turn
means that Eq.~(\ref{Stab11}) also holds, i.e.,
the heat capacity $C_A$ is positive,
$C_A\geqslant0$. Specifically we found that
for free $M$ and fixed $A$,
\begin{equation}\label{sta00}
{\rm Stability\, always\, }\,
\quad {\rm for}\; -1<a \leqslant-\frac{1}{d-2}\,,
\end{equation}
and Eq.~(\ref{quad1}) is satisfied when
${0\leqslant k \leqslant k_1}$ for ${-1/(d-2)<a<\infty}$, i.e., 
\begin{equation}\label{stabk0}
{\rm Stability\, when\,}0\leqslant k \leqslant k_1,\,\quad {\rm for}\;
-\frac{1}{d-2}<a<\infty\,,
\end{equation}
with 
\begin{equation}\label{k0}
k_1= \sqrt{\frac{d-3}{d-1 + 2a(d-2) }}\,.
\end{equation}
Note anew
that $0<k_1\leqslant1$ in this case, i.e., for ${-1/(d-2)<a}$.
Using Eq.~(\ref{k})
we can put condition Eq.~(\ref{stabk0}) in terms of $R/r_+$, 
\begin{equation}\label{r+R1}
1\leqslant \frac{R}{r_+}\leqslant \frac{1}{(1-k_1^2)^{\frac{1}{d-3}}}\,,
\end{equation}
for $-1/(d-2)<a<\infty$.
Changing from $r_+$ to $M$ through
Eqs.~(\ref{rplus}) and~(\ref{E1})
we obtain
that 
the thin shell's
radius is bounded from above as
$\frac{2R^{d-3}}{{\gamma}M} \leqslant \frac{1}{1 - k_1}$.

The case $a=0$ is of particular interest
as the inverse temperature has the inverse Hawking temperature
form, it is proportional to $r_+$, see Eq.~(\ref{C1}).
Putting $a=0$ in
$k_1$, see Eq.~(\ref{k0}), we get
$k_1=\sqrt\frac{d-3}{d-1}$ and the
stability is given then by  Eq.~(\ref{stabk0}).
One can solve for $\frac{R}{r_+}$, see also Eq.~(\ref{r+R1}), to
obtain
$1\leqslant\frac{R}{r_+}\leqslant\left(\frac{d-1}{2}\right)^{\frac{1}{d-3}}$.
Looking at Eq.~(\ref{rph}) we see this is
\begin{equation}\label{shella=0tab}
r_+\leqslant R\leqslant r_{\rm ph}\,. 
\end{equation}
I.e., the shell is thermodynamically stable if its radius $R$ is in
between the gravitational radius $r_+$ and the photon sphere radius
$r_{\rm ph}$.  For $d=4$ Eq.~(\ref{shella=0tab}) is ${r_+ \leqslant
R\leqslant \frac32 r_+}$, and putting $r_+=2m$ one gets $2m \leqslant
R\leqslant 3m$.  This
is a striking outcome as it reminds of York's result for the thermal
stability of the $d=4$ black hole in the canonical ensemble
\cite{york1}. York's approach implies that for a Scharzschild black
hole in a heat reservoir of fixed radius $R$ at temperature $T$, i.e.,
in a canonical ensemble, the heat capacity of the black hole system is
positive only if $2m \leqslant R\leqslant 3m$, so in this range the
system is stable. Our result says that for $a=0$, the heat capacity is
positive if the shell is in the range $2m \leqslant R\leqslant
3m$. The shell's heat capacity is measured for $A$ fixed, i.e., $R$
fixed, and the shell itself acts as a heat reservoir. The two systems
have thus similarities but are different.  One is a black hole in a
heat reservoir at temperature $T$, the other a massive shell at
temperature $T$, one has a Schwarzschild interior to the heat
reservoir, the other a flat interior to the shell.  This is an
unexpected result and hints that what is important for thermodynamic
stability is the place of the shell alone, being it a heat reservoir
massless shell or a massive shell.

\begin{figure}[h!]
\centering
\includegraphics[width=0.4\textwidth]{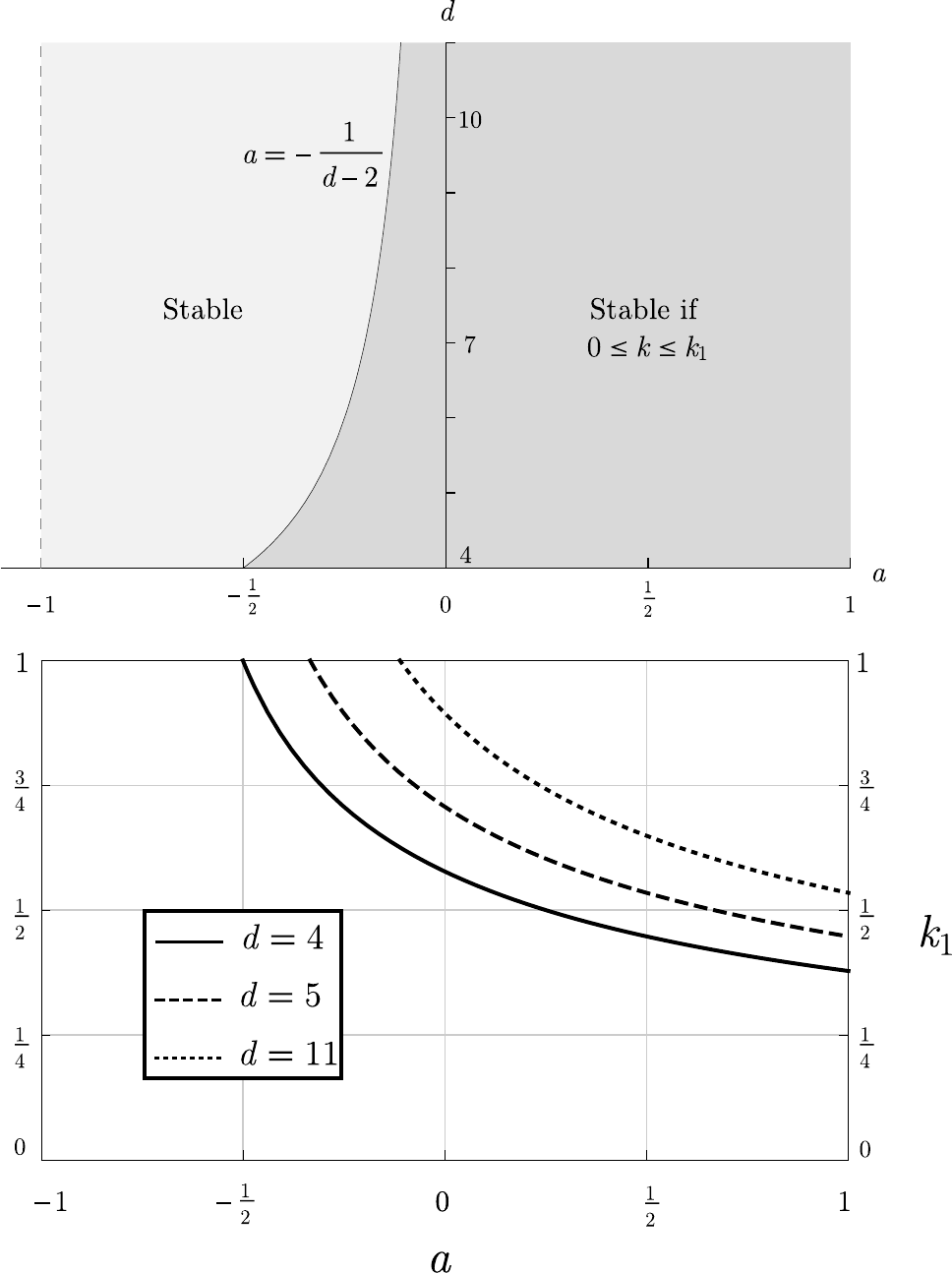}
\caption{Intrinsic stability of the shell for free proper mass $M$ and
at fixed area $A$, i.e., fixed radius $R$, given in Eqs.~(\ref{sta00})
and~(\ref{stabk0}), for a self-gravitating shell with a power-law
temperature equation of state.  Stability at fixed area $A$ is
equivalent to have a heat capacity $C_A$ obeying $C_A\geqslant0$.
Top: The stability regions in the parameter space given by the
equation of state exponent $a$ versus the number of dimensions $d$, an
integer number with $d\geqslant4$, are displayed.  In the region
${a \leqslant -\frac1{d-2}}$, the shell is always stable.
In the region
${a>-\frac1{d-2}}$, the shell is only stable
if ${0\leqslant k \leqslant k_1}$.
Bottom: Plot of $k_1$ given in Eq.~(\ref{k0}) as a function of $a$ for
three different values of $d$, $d=4,5,11$. The shell is stable if $k$
is below the respective $k_1$.  See text
for details.}
\label{pspacek1}
\end{figure}

In Fig.~\ref{pspacek1}, we plot the stability regions in the parameter
space given by the equation of state exponent $a$ versus the number of
dimensions $d$, an integer number with $d\geqslant4$.  Adjacently, we
also plot the quantity $k_1$ given in Eq.~(\ref{k0}) in terms of the
equation of state exponent $a$ for three different dimensions $d$,
$d=4,5,11$. This makes it easier to follow the stability parameters.

\subsubsection{Stability for fixed proper mass
$M$ and for free area $A$ }\label{fixedM}

Condition Eq.~(\ref{Stab2}) is the stability condition if  the 
proper mass $M$ of the shell is held fixed and the
area $A$ is free to change.
Equation~(\ref{S1}) together with
Eqs.~(\ref{rplus}), (\ref{ar1}),
and~(\ref{E1})
yields
$\left( \frac{\partial^2S}{\partial A^2} \right)_M =
\frac{(1+a)S(M,A)}{A^2(d-2)(1-k)}
\left[ (1-k)(2+a)(d-2)-2(2d-5)  \right]$.
Thus, Eq.~(\ref{Stab2}) gives
\begin{equation}\label{quad2}
	-k(d-2)(a+2)-2(2d-5)+(d-2)(a+2)\leqslant0\,.
\end{equation}
The solution for Eq.~(\ref{quad2}) is $k \geqslant k_2$ where $k_2 =
\frac{a-2\frac{d-3}{d-2}}{a+2}$.  Recalling from Eq.~(\ref{a>-1}) that
$a>-1$, we note that $k_2\leqslant0$ for $-1<a\leqslant
2(d-3)/(d-2)$ and,
since $0\leqslant k \leqslant1$, Eq.~(\ref{quad2}) is always satisfied. For
$a > 2(d-3)/(d-2)$ we have $0<k_2<1$, so Eq.~(\ref{quad2}) is
satisfied if $k_2 \leqslant k \leqslant 1$.

\begin{figure}[h!]
\centering
\includegraphics[width=0.4\textwidth]{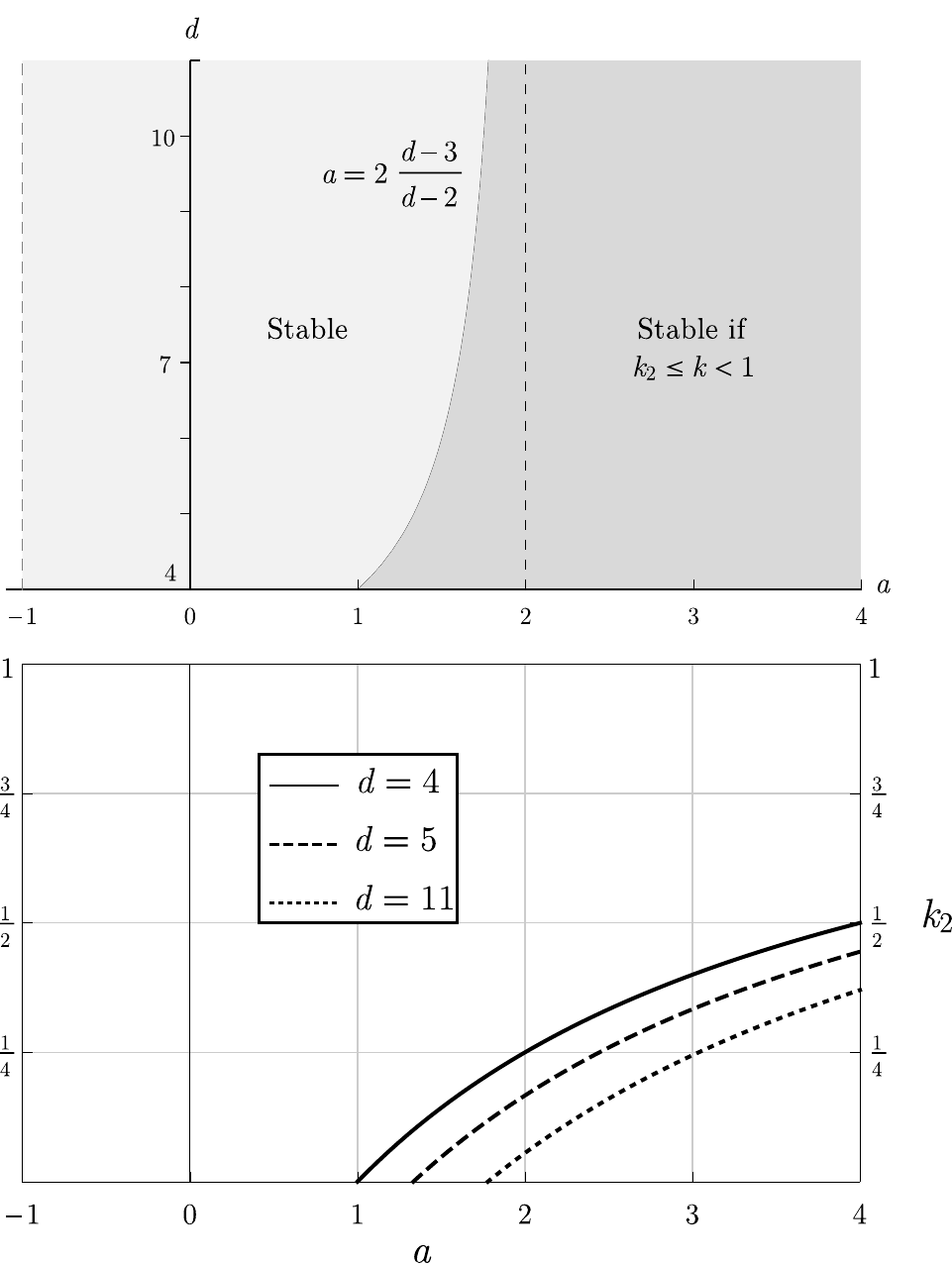}
\caption{Intrinsic stability for free area $A$ and fixed proper mass
$M$, given in Eqs.~(\ref{sta01}) and~(\ref{stabk01}), for a
self-gravitating shell with a power-law temperature equation of state.
Top: The stability regions in the parameter space given by the
equation of state exponent $a$ versus the number of dimensions $d$, an
integer number with $d\geqslant4$, are displayed.  In the region ${a \leqslant 2
\frac{d-3}{d-2}}$, the shell is always stable.  In the region ${a>2
 \frac{d-3}{d-2}}$, the shell is only stable if ${k_2\leqslant k \leqslant 1}$.
Bottom: Plot of $k_2$ as a function of $a$ for three different values
of $d$, $d=4,5,11$.  The shell is stable if $k$ lies above the
respective $k_2$. See text for details.}
\label{pspacek2}
\end{figure}

So in brief, stability for fixed $M$ and free $A$ means that
Eq.~(\ref{Stab2}) holds.  Specifically we found that for
fixed $M$ and free $A$,
\begin{equation}\label{sta01}
{\rm Stability\, always\, }\,
\quad {\rm for}\; -1<a\leqslant 2\frac{d-3}{d-2}\,,
\end{equation}
and Eq.~(\ref{quad1}) is satisfied when
$k_2 \leqslant k \leqslant 1$
for
$a>2(d-3)/(d-2) $, i.e., 
\begin{equation}\label{stabk01}
{\rm Stability\, when\,}k_2 \leqslant k \leqslant 1\,\quad {\rm for}\;
a>2\frac{d-3}{d-2}\,,
\end{equation}
with 
\begin{equation}\label{k11}
k_2= \frac{a-2\frac{d-3}{d-2}}{a+2}\,.
\end{equation}
Using Eq.~(\ref{k}) we can write 
condition Eq.~(\ref{stabk01}) in terms of $R/r_+$, 
\begin{equation}\label{r+R2}
\frac{R}{r_+} \geqslant \frac{1}{(1-k_2^2)^{\frac{1}{d-3}}}\,,
\end{equation}
for $a>2(d-2)/(d-2)$.
Changing from $r_+$ to $M$ through
Eqs.~(\ref{rplus}) and~(\ref{E1})
we obtain
that 
the thin shell's
radius is bounded from below as
$\frac{2R^{d-3}}{{\gamma} M} \geqslant \frac{1}{1 - k_2}$.

For the particularly interesting case
$a=0$ we see from 
Eq.~(\ref{sta01})
that the $a=0$  shell is thermodynamically stable
for any radius, as
the condition is independent of it.

In Fig.~\ref{pspacek2}, we plot the stability regions in the parameter
space given by the equation of state exponent $a$ versus the number of
dimensions $d$, an integer number with $d\geqslant4$.
Adjacently, we also plot the quantity $k_2$ given in
Eq.~(\ref{k11}) in terms of the equation of state exponent $a$ for
three different dimensions $d$, $d=4,5,11$. This makes it easier to
follow the stability parameters.

\subsubsection{Stability for free proper mass $M$ and free
area $A$}\label{stabilityMA}

In the case $M$ and $A$ are free, 
condition Eq.~(\ref{Stab3}) also counts and one needs
to check it. It will be seen
that Eq.~(\ref{Stab3}) yields 
the most stringent conditions between the three conditions.
Equation~(\ref{S1}) together with
Eqs.~(\ref{rplus}), (\ref{ar1}),
and~(\ref{E1})
yields
$
\frac{\partial^2 S}{\partial M \partial A}  =
\frac{2(1+a)S(M,A)}{M(d-3)(1-k)} \left[ -(1-k)(1+a)(d-2) +
2d-5+a(d-2) \right]$.
Thus, Eq.~(\ref{Stab3}) gives
\begin{align}
&k^2(2+a(d-1))+\frac{2k(d-3)(1+a(d-2))}{d-2} +\nonumber \\
& +a(d-3) \leqslant 0 \label{quadratic2}\,.
\end{align}
For ${-1<a\leqslant-2/(d-1)}$, the inequality is always satisfied
by any $0\leqslant k \leqslant1$.
For $a>-2/(d-1)$, the inequality is satisfied by ${k_{3-}\leqslant k \leqslant
k_3}$, where $k_{3-}$ and $k_3$ are the roots in
Eq.~(\ref{quadratic2}).
Since ${k_{3-}<0}$, it can be discarded, and
the inequality is satisfied by any $0\leqslant k \leqslant k_3$, where
$
k_3 = -\frac{(d-3)(1+a(d-2))}{(2+a(d-1))(d-2)} 
+ \frac{\sqrt{(d-3)(d-3-2a(1+a(d-2))(d-2)) }}{(2+a(d-1))(d-2)}
$.
However, for ${-2/(d-1)<a \leqslant -1/(d-2)}$ note that ${k_3 \geqslant 1}$, so
that the inequality is satisfied by any ${0\leqslant k \leqslant 1}$. For
${-1/(d-2)<a \leqslant 0}$ we have that ${0\leqslant k_3 < 1}$, so that the
inequality is satisfied by ${0\leqslant k \leqslant k_3}$. For $a>0$ note that
${k_3<0}$, so the inequality cannot be satisfied.

\begin{figure}[h!]
\centering
\includegraphics[width=0.4\textwidth]{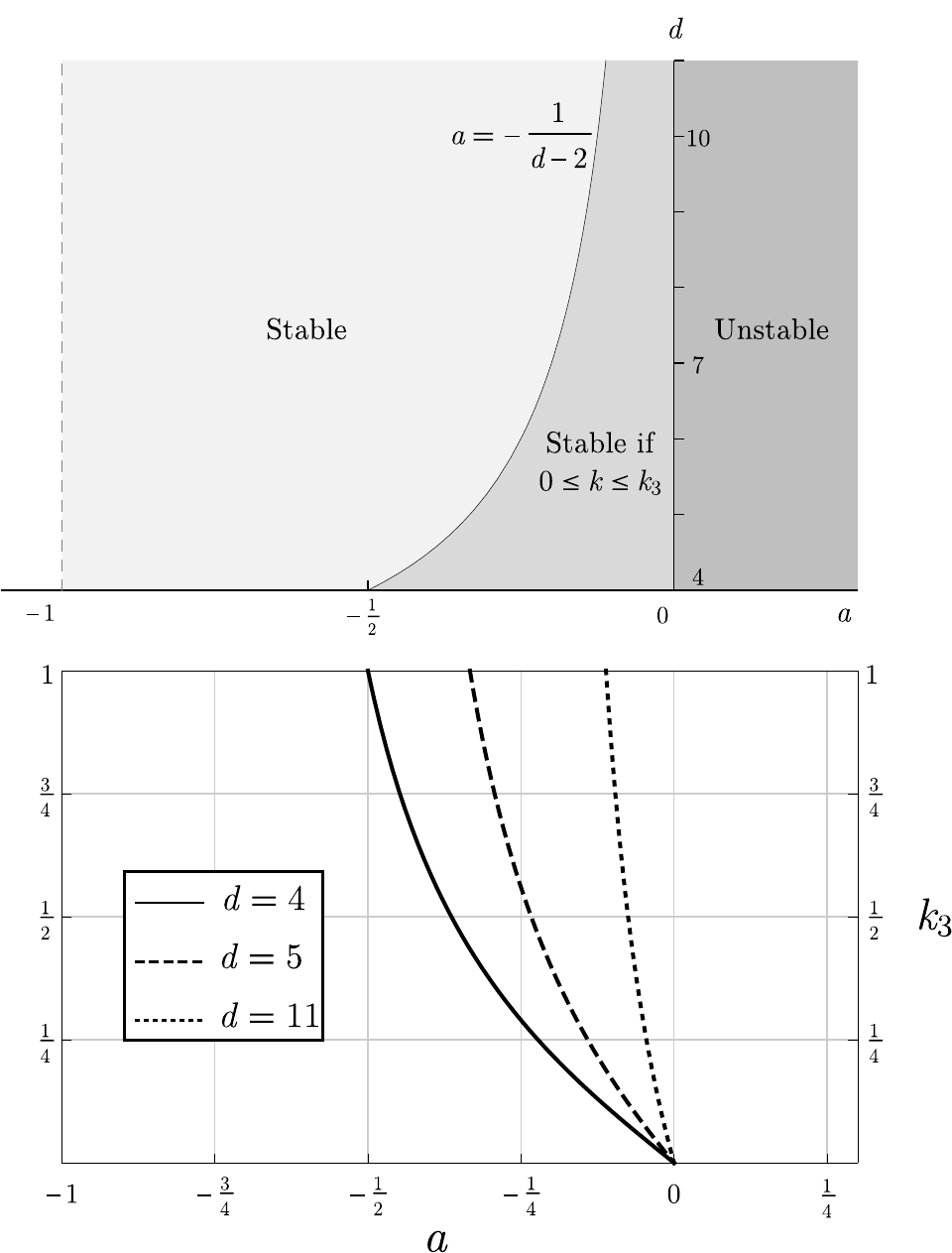}
\caption{Intrinsic stability for free proper mass $M$ and free area
$A$, given in Eqs.~(\ref{sta22}),~(\ref{sta222}), and~(\ref{sta2222}),
for a self-gravitating shell with a power-law temperature equation of
state. Top:  The stability regions in the parameter space given by the
equation of state exponent $a$ versus the number of dimensions $d$,
an integer number with $d\geqslant4$, are
displayed.  In the region ${-1 < a \leqslant -\frac{1}{d-2}}$, the shell is
always stable.  In the region ${-\frac{1}{d-2} < a \leqslant 0}$, the shell is only
stable if $0 \leqslant k \leqslant k_3$.  In the region $a>0$, the shell never
satisfies the thermodynamic stability criterion. Bottom: Plot of $k_3$
as a function of $a$ for three different values of $d$, $d=4,5,11$.
The shell is stable if $k$ lies above the respective $k_3$.  In the
region $a>0$ one has $k_3<0$
and there is no stability. See text for details.}
\label{pspacek3}
\end{figure}

So in brief,  stability for free $M$ and free $A$ means that
Eq.~(\ref{Stab3}) holds.  Specifically we found that for
free $M$ and free $A$, the solutions have
\begin{equation}\label{sta22}
{\rm Stability\, always\, }\,
\quad {\rm for}\;	 -1 < a \leqslant -\frac{1}{d-2}\,,
\end{equation}
and 
Eq.~(\ref{quadratic2}) is satisfied when
$0 \leqslant k \leqslant k_3$
for
${ -1/(d-2) < a \leqslant 0}$, i.e., 
\begin{equation}\label{sta222}
{\rm Stability\, when\,}0 \leqslant k \leqslant k_3\,\quad {\rm for}\;
-\frac{1}{d-2} < a \leqslant 0
\end{equation}
with
\begin{align}\label{k3}
k_3 = &-\frac{(d-3)(1+a(d-2))}{(2+a(d-1))(d-2)} + \nonumber \\
+ &\frac{\sqrt{(d-3)(d-3-2a(1+a(d-2))(d-2)) }}{(2+a(d-1))(d-2)}1,
\end{align}
and 
for $a>0$, there are no thermodynamically stable configurations,
i.e.,
\begin{equation}\label{sta2222}
{\rm No\,\, stability\,}\,\quad {\rm for}\;a>0\,.
\end{equation}

Using Eq.~(\ref{k}), Eq.~(\ref{sta222}) can be written in
terms of $R/r_+$ as 
\begin{equation}\label{r+R3}
1\leqslant \frac{R}{r_+}\leqslant \frac{1}{(1-k_3^2)^{\frac{1}{d-3}}}
\,.
\end{equation}
for ${-1/(d-2) < a \leqslant 0}$.
Changing from $r_+$ to $M$ through
Eqs.~(\ref{rplus}) and~(\ref{E1})
we obtain
that 
the thin shell's
radius is bounded from above as
$\frac{2R^{d-3}}{{\gamma} M} \leqslant \frac{1}{1-k_3}$.

For the particularly interesting $a=0$ case 
the stability condition is given in Eq.~(\ref{sta222}).
It involves
the quantity $k_3$ given in Eq.~(\ref{k3}) which for
$a=0$ gives $k_3=0$. 
This means that to be thermodynamic stable
under these perturbations the radius of the shell $R$
obeys $R=r_+$. For $a=0$ this is the only
thermodynamic stable case.

In Fig.~\ref{pspacek3}, we plot the stability regions in the parameter
space given by the equation of state exponent $a$ versus the number of
dimensions $d$, an integer number with $d\geqslant4$.
Adjacently, we also plot the quantity $k_3$ given in
Eq.~(\ref{k3}) in terms of the equation of state exponent $a$ for
three different dimensions $d$, $d=4,5,11$. This makes it easier to
follow the stability parameters.

\subsubsection{Summary of the  stability analysis: All three cases
together}\label{stabilityfinal}

Collecting the results
for the stability of a self-gravitating shell,
we see that the third
condition is the stricter one for stability. Indeed:
(i) Eqs.~(\ref{sta00}) and~(\ref{sta22}) 
give the same result and are stricter than~(\ref{sta01})
in the range of $a$.
(ii) In the range $-1/(d-2) < a \leqslant 0$ of 
Eq.~(\ref{sta222}), $k_3<k_1$ always, so 
Eq.~(\ref{sta222}) makes Eq.~(\ref{stabk0}) spurious
in this range of $a$.
(iii) In the range $a>0$
of 
Eq.~(\ref{sta2222}) all solutions are unstable, so 
Eqs.~(\ref{stabk0}) and~(\ref{stabk01}) are irrelevant in this range.
Thus, 
Eqs.~(\ref{sta22})-(\ref{sta2222}) are the ones necessary and sufficient
for intrinsic local thermodynamic
stability. Nonetheless, Eqs.~(\ref{sta00}) and~(\ref{stabk0})
are valid for thermodynamic
 stability at fixed $A$
and Eqs.~(\ref{sta01}) and~(\ref{stabk01})
are valid for thermodynamic stability at fixed $M$.

\subsection{Euler relation}

From the entropy $S$ in Eq.~(\ref{S1}), and using the expression for the
ADM mass $m$ in terms of the proper mass
$M$ given in Eq.~(\ref{E1}),
one finds that
\begin{equation}
\frac{{\gamma} M}{2R^{d-3}} = 1-\left[ \frac{1}{R^{d-3}}
\left( \frac{4S}{\eta \,\Omega_{d-2}} \right)^\frac{d-3}{(a+1)(d-2)}
\right]^{\frac12}\,.
\end{equation}
Applying Euler's
theorem on homogeneous functions \cite{callen} to $M$, which
is homogeneous of degree
$\frac{d-3}{(a+1)(d-2)}$ in $S$ and 
$\frac{d-3}{d-2}$ in $A$,
yields the Euler relation for this system,
\begin{equation}\label{euler1}
M= \frac{(a+1)(d-2)}{d-3} \, TS - \frac{d-2}{d-3} \, pA\,.
\end{equation}
From the presence of the free parameter $a$, we see that
the Euler relation is
dependent on the equation of state for the temperature.

The scaling laws for the self-gravitating shell are $M\to\lambda M$,
$S\to\lambda^{ \frac{(a+1)(d-2)}{d-3}} S$, and $A\to\lambda^{
\frac{d-2}{d-3}} A$.  This makes sense. Indeed, under this rescaling
one has from Eq.~(\ref{C1}) $T\to\lambda^{-1-a(d-2)} T$ and from
Eq.~(\ref{p11}) $p\to\lambda^{- \frac{1}{d-3}}p $, altogether make
$M\to\lambda M$ in Eq.~(\ref{euler1}).

Taking the differential of Euler's relation Eq.~(\ref{euler1}) and
taking into account the first law Eq.~(\ref{1T}) one obtains the
Gibbs-Duhem relation for this system
\begin{align} 
(a+1)(d-2)SdT + (1+a(d-2))dM& \nonumber \\
+ a(d-2)pdA - (d-2)Adp &= 0.
\end{align}

An interesting case is $a=0$. 
So let us put $a=0$ in the equation of state for the shell's
temperature. Then 
from Eq.~(\ref{euler1}) we find then that the Euler relation
for such a shell reads
$
M= \frac{d-2}{d-3} \, TS - \frac{d-2}{d-3} \, pA
$,
and the shell's proper mass is an homogeneous function of degree
$\frac{d-3}{d-2}$ in $S$ and $A$.
The scaling laws for the self-gravitating  shell in this
$a=0$ case 
are $M\to\lambda M$,
$S\to\lambda^{ \frac{d-2}{d-3}} S$, and 
$A\to\lambda^{ \frac{d-2}{d-3}} A$.
Taking the differential
of Euler's relation Eq.~(\ref{euler1}) and
taking into account the first law 
Eq.~(\ref{1T}) 
one obtains the
Gibbs-Duhem relation for this system
$
(d-2)SdT + dM
 - (d-2)Adp = 0
$.

\subsection{Other topics on entropy}

\subsubsection{Bekenstein 
entropy bound for the $d$-dimensional 
shell}\label{bbounds}



The Bekenstein bound relates the entropy
and the energy of a system. 
We follow the argument
presented by Bekenstein for four dimensions
in~\cite{Bekenstein}
and turn it into a $d$-dimensional bound.
Given a $d$-dimensional spherical object with
energy $E$, size $l$ and entropy $S_E$, and a black hole
with horizon radius and area $r_+$ and $A_+$,
respectively, and entropy $S_+=\frac14 A_+$, 
one has that the initial entropy $S_{\rm i}$
of the system 
is ${S_{\rm i}= \frac14 {A_+} + S_E}$.
If the object is swallowed by the black hole,
this will grow by an area $\Delta A_+$,
so the
final entropy is 
$S_{\rm f} = \frac14 \left( A_+ +\Delta A_+\right)$.
From
$r_+^{d-3}={\gamma} \, m$, see Eq.~(\ref{rplus}),
and $A_+=\Omega_{d-2}r_+^{d-2}$, see Eq.~(\ref{gar1}),
we get
$\Delta A_+ = {\gamma} \,\Omega_{d-2}
\left( \frac{d-2}{d-3} \right) E\,r_+$,
where we have naturally put $E=\Delta m$.
For the generalized
second law of thermodynamics to hold, 
one must have $S_{\rm i}\leqslant S_{\rm f}$, so 
then $S_E \leqslant\frac{{\gamma} \Omega_{d-2}}4 \left(
\frac{d-2}{d-3} \right)E r_+$.
If $l$ is not small compared to $r_+$ then a
bound like
$S_E < \alpha \frac{{\gamma} \Omega_{d-2}}4 \left( \frac{d-2}{d-3}
\right)E l$ must exist, for some value of $\alpha$ which cannot be
set by this 
argument. We choose $\alpha$ as
$\alpha=\frac{d-3}{d-2}$ for reasons given below.
Using ${\gamma}\,\Omega_{d-2}=\frac{16\pi}{d-2}$,
see Eq.~(\ref{gamma1}),
the bound is 
\begin{equation}\label{bekbound}
S_E \leqslant \frac{4\pi}{d-2}\, E l\,.
\end{equation}
Now, although the bound was here suggested through a
definite example
involving matter and a black holes,
Eq.~(\ref{bekbound}) is assumed to be valid
for all matter in all kinds of situations and is 
called the Bekenstein bound. In particular it can be
applied to the self-gravitating shells
we have been considering.

Let us suppose a self-gravitating
shell with energy $E$ and typical length
$l$. In the shell case the quantity $E$ can
have two interpretations. It can 
be interpreted either
as the rest mass
$M$ of the shell, $E=M$, or
as the ADM mass of the spacetime, $E=m$. The length
$l$
can be put equal to the radius of the system $l=R$. 

For $E=M$ the bound Eq.~(\ref{bekbound}) 
for the entropy $S$ of the shell is
\begin{equation}\label{bekbound1}
S \leqslant \frac{4\pi}{d-2}\, MR\,.
\end{equation}
Using  
Eq.~(\ref{M1}) together with 
Eq.~(\ref{gamma1}) this can be put as
$S \leqslant \frac12 \Omega_{d-2} (1-k) R^{d-2}$.

For $E=m$ the bound Eq.~(\ref{bekbound}) 
for the entropy $S$ of the shell is 
\begin{equation}\label{bekbound2}
S \leqslant \frac{4\pi}{d-2}\, mR\,.
\end{equation}
Using  
Eq.~(\ref{rplus}) together with 
Eq.~(\ref{k}) this yields
$
S < \frac14  \Omega_{d-2}
(1-k^2) R^{d-2}$.

Which bound shall one choose, if the one given
in Eq.~(\ref{bekbound1})
or the one given
in Eq.~(\ref{bekbound2}), cannot be settled
by this analysis.

\subsubsection{Holographic entropy bound for the $d$-dimensional 
shell}\label{holdbound}


The holographic entropy bound~\cite{thooft,Susskind}
claims that in a full developed
theory of quantum gravity the  entropy $S_A$ 
in a region enclosed by an area $A$ is always less or equal
to $A/4$ in Planck units, 
\begin{equation}\label{Smax}
S_A\leqslant\frac{A}{4}\,.
\end{equation}
One insight for the bound came
from the gravitational collapse of a star of area $A$ and the
entropy law governing black holes~\cite{Susskind}.
For a black hole the entropy is precisely equal to one quarter
of its horizon area, so black holes saturate the inequality
Eq.~(\ref{Smax}).
It is further conjectured that the bound also holds for
higher $d$-dimensions, with the area $A$ being a $d-2$
surface $A$ enclosing a $d-1$ volume~\cite{Hod3}.

In the case in hand we have a thermodynamic self-gravitating shell
with a particular equation of state, 
Eq.~(\ref{C1}).
It is thus relevant to
know whether the holographic entropy bound is automatically
satisfied or if both the junction and stability conditions still
allow for configurations whose entropy exceeds the bound.
Since the holographic bound 
questions the feasibility of a physical system
that exceeds it, it  is relevant to see
how it works for shells.

For the shell's area Eq.~(\ref{ar1}) and the shell's entropy
Eq.~(\ref{S}), the entropy bound Eq.~(\ref{Smax}) is satisfied if
$
\eta\,r_+^{(a+1)(d-2)}\leqslant R^{(d-2)}
$, i.e., 
\begin{equation}\label{Bbound1}
\frac{R}{r_+} \geqslant \eta^{\frac{1}{d-2}}\,r_+^a\,.
\end{equation}
Given $a$ and $\eta$  the bound is irrelevant if
$ \eta^{\frac{1}{d-2}} r_+^a \leqslant 1 $,
since in this case $\frac{R}{r_+}\geqslant 1$, see
Eq.~(\ref{Rr+b}), puts a stronger limit
on $R$ and so the bound is
always obeyed. For instance for $-1<a\leqslant 0$ and
${0< \eta\leqslant1}$ the bound is irrelevant.
We have found that solutions where $-1<a\leqslant 0$
are thermodynamically
stable solutions for all dimensions $d$. It is 
notable that  stable solutions for all $d$  
obey automatically the holographic entropy bound. 
If 
$
\eta^{\frac{1}{d-2}}r_+^a > 1
$, then only those configurations whose $R$ obeys
Eq.~(\ref{Bbound1}) are the ones that
satisfy the bound.

\subsubsection{Entropy of the shell for large $d$}

When generalizing a physical system to higher dimensions
one should understand how the physical quantities,
in particular the entropy, change with the dimension.
In particular, for the entropy
this might have some
implications on whether or not the system stays within
one ot both entropic
bounds. In this
connection, the $d\to\infty$
limit is useful and interesting.
We will take the limit $d\rightarrow \infty$, and see
how the entropy of a
self-gravitating
thin shell acts in response.
To do so, we
write the solid angle given in Eq.~(\ref{O}) in the following way, using the
Stirling approximation,
$
\Omega_{d-2} = \sqrt{\frac{2}{e}} \left( \frac{2\pi e}{d-3}
\right)^{\frac{d-2}{2}}\left(1+{O}\left( \frac{1}{d} \right) \right).
$
Although the approximation works better as $d \rightarrow \infty$, it
 is also a good fit for any $d\geqslant4$.  For the shell's entropy
given in Eq.~(\ref{S1}) we find
\begin{equation}\label{sdlarge}
S = \left(\frac{2\pi e}{d}\right)^{d/2} \eta\, r_+^{d(a+1)}\,.
\end{equation}
Clearly we have that $S\rightarrow 0$ as $d$ grows, and this is
because the solid angle converges very quickly to zero, with
$1/d^{d/2}$.
Instead of setting $\eta$ as a constant, we could
include the $1/d^{d/2}$ factor into ${\eta_d \equiv \eta / \,d^{d/2}}$
and set it as our problem's constant. But we will not do this here.
One can also see how the large $d$ limit affects the distance to
the holographic bound. Computing the ratio between the two, the solid
angle terms cancel out, and we find
$
\frac{S}{A/4} = \eta
\left( \frac{r_+^{a+1}}{R} \right)^{d}  
$.
Now, as mentioned previously if ${a\leqslant0}$ and
${\eta \leqslant 1}$ the bound is always satisfied, and
we see that, as $d$ increases, the shell's entropy
will distance itself farther from the bound, i.e.,
in this case it holds that
$
\frac{S}{A/4}  \leqslant1
$.

We can additionally study the behavior of the
other physical quantities of the shell.
Since $(1-k)$ goes to zero with ${\left( \frac{r_+}{R}\right)^{d}}$,
see Eq.~(\ref{k}),
both the shell's rest mass $M$ and pressure $p$ go to zero,
as one can check in Eqs.~(\ref{M1}) and (\ref{p11}).
Because $k\rightarrow 1$, the temperature is $T=1/b$ and its
behavior at the large $d$ limit depends on the sign of
the equation of state exponent $a$, as 
can be seen in Eq.~(\ref{C1}).
For $a\leqslant 0$,  the temperature $T$ diverges with $d r_+^{-a d}$, 
whereas for $a>0$ $T$  goes to zero in the $d\to\infty$ limit.

\section{Black holes in $d$-dimensions: Entropy,
local thermodynamic stability, Smarr formula, 
entropy bounds, and large $d$}
\label{bh1}

\subsection{Black hole equation of state and entropy}

We are now interested in studying black hole properties in
$d$-dimensions using the results from thin shells. For that we take
the $d$-dimensional shell to its gravitational radius $R\rightarrow
r_+$, i.e., we take the quasiblack hole limit \cite{lz1,lz2}.
At this quasiblack hole stage the exterior spacetime to the
shell is that of 
$d$-dimensional Schwarzschild black hole, i.e., Tangherlini black
hole.

To do this note that one possible equation of state for the
temperature of the shell is the Hawking temperature $T_+$ given by
$T_+= \frac{d-3}{4\pi} \frac{1}{r_+}$ \cite{kanti}, i.e., the inverse
temperature is $b_+=\frac{4\pi}{d-3} r_+$. From Eq.~(\ref{C1}) for the
inverse temperature $b(r_+)$ of the shell at infinity one sees that
putting $a=0$ and $\eta=1$ one recovers precisely $b_+$. In this case
from Eq.~(\ref{S1}) the entropy $S$ of the shell with radius $R$ is
$S= \frac14 A_+$. We now can take the limit and send the radius $R$ of
the shell to its own gravitational radius $r_+$, $R\rightarrow r_+$.

Before we do that
we note that
when performing the quasistatic collapse of the shell into $r_+$,
the only reasonable equation of state for the
inverse temperature is indeed
$b_+$.  The analysis we have been doing demands
thermal equilibrium so that we can safely
use the first law of thermodynamics
Eq.~(\ref{1T}).  If then we take into account that quantum fields are
present just outside the shell at its own gravitational radius
$R=r_+$, the shell's inverse temperature must be the black hole
inverse temperature, so $b(r_+)$ in Eq.~(\ref{C1}) must have the
expression $b(r_+)=b_+=\frac{4\pi}{d-3} r_+$ in order to have
equilibrium.  Moreover, it has been shown in some particular
instance~\cite{loranz} that the thermal energy-momentum tensor
$T^a_{\,\,b}$ for a field at temperature $T_{\rm field}$ is of the
form~\cite{loranz} $T^a_{\,\,b}=
\frac{T_{\rm field}^4-T^4_+}{g_{00}^2}f^a_{\,\,b}$, for some tensor
$f^a_{\,\,b}$ finite at the horizon, with $g_{00}$ being the time-time
metric component of the static spacetime. Assuming this is also valid
in $d$-dimensions we see that since $g_{00}$ is zero at $R=r_+$, $T^a_{\,\,b}$
diverges unless the temperature of the field $T_{\rm field}$ is the Hawking
temperature $T_+$, $T_{\rm field}=T_+$, and so $b_{\rm field}=b_+$.

So when one collapses the shell quasistatically
into a black hole, i.e., $R=r_+$, Eq.~(\ref{C1})
must take the form 
\begin{equation}\label{thawk}
b_+(r_+)=\frac{4\pi}{d-3} r_+\,. 
\end{equation}
Then the entropy from Eq.~(\ref{S1})
is
\begin{equation}\label{SBH}
S_+=\frac14 
A_+\,. 
\end{equation}
This is the 
Bekenstein-Hawking entropy in $d$-dimensions,
obtained here from the self-gravitating shell
formalism with the input of
the Hawking temperature.

\subsection{Black hole intrinsic thermodynamic
stability}~\label{IntStability3}

For a black hole $a=0$ and $\eta=1$.
In the black hole limit, we take in addition
$R\rightarrow r_+$ implying $k\rightarrow 0$. 
The stability equations
we are interested are given in
Eqs.~(\ref{stabk0}),~(\ref{sta01}), and~(\ref{sta222}).
We first take $a=0$ in the stability conditions
and see the properties for the shell with this $a$.
Then we take the black hole limit $R\rightarrow r_+$
and discuss the features in this case.

For fixed shell's area $A$, $a=0$, and $R\rightarrow r_+$,
thermodynamic stability
is taken from Eq.~(\ref{stabk0}) or Eq.~(\ref{shella=0tab}).
Clearly one finds
that the shell is
thermodynamically stable at the gravitational radius for $d\geqslant 4$.
Indeed, when the shell with $a=0$ is at its gravitational
or horizon radius, i.e., $k=0$, it 
satisfies marginally the intrinsic thermodynamic
stability criterion Eq.~(\ref{stabk0}).  This is
because the heat capacity
$C_A = T^{-2}\left( \frac{\partial^2 S}{\partial M^2}
\right)_A^{-1}$
goes to zero with $T^{-2}$
in this
limit. Since $C_A$ is also defined as 
$C_A = \left( \frac{\partial M}{\partial T}
\right)_A$, $C_A=0$ means that the mass of the shell
cannot be altered by any change
on the infinitely high temperature.
In this limit
we cannot increase the
mass $M$ of the shell
with $A=A_+$, i.e., $R=r_+$, constant,
since from Eq.~(\ref{M1}) in this limit one has 
$\frac{{\gamma}}2 M = r_+^{d-3}$. So,
to change $M$ one has to change 
the radius $R=r_+$.
Moreover,
York's results for black holes in a canonical ensemble
\cite{york1} imply
that when the heat reservoir is placed at $R=r_+$ the black hole is
thermodynamically marginally
stable.  The two results are indeed the same 
as
the two situations deal with
the same black hole,
namely,
a black hole in a heat reservoir at its 
horizon at temperature $T$.
So, 
York's heat reservoir  at the black hole
horizon and the massive shell at the gravitational
radius are the same thing, and York's criterion for thermodynamic
stability is precisely reproduced.

For fixed proper mass $M$, $a=0$, and $R\rightarrow r_+$
thermodynamic stability comes from  Eq.~(\ref{sta01})
and we have seen that
the $a=0$ shell is stable under this condition
for all radii in particular for $R=r_+$.

For free $A$ and $M$, $a=0$, and $R\rightarrow r_+$
thermodynamic stability comes from  Eq.~(\ref{sta222}).
Since in this case $k_3=0$
as we have seen, the only stable case is
precisely $R=r_+$, i.e.,
the black hole is stable.

\subsection{Smarr relation}
\label{smarr}

For a black hole $a=0$ and 
the black hole  Euler relation
has to be taken from Eq.~(\ref{euler1}) putting $a=0$.
In the black hole limit, we take in addition
$R\rightarrow r_+$ implying $k\rightarrow 0$. 
Since both the temperature $T$ and pressure $p$ go with $k^{-1}$,
one has $kM=\frac{d-2}{d-3} \, T_+S_+- \frac{d-2}{d-3} \, p_+A_+$,
where $T_+=1/b_+$ is the Hawking temperature with 
$b_+$ given in Eq.~(\ref{thawk}), 
$S_+$ is the Bekenstein-Hawking entropy 
given in Eq.~(\ref{SBH}), $p_+$ is the redshifted
pressure $p_+=pk$ with $p$
given in Eq.~(\ref{p11}), and 
$A_+$ the horizon area 
given in Eq.~(\ref{gar1}). 
So, since $k\to0$ in this limit,
this translates into
$0= \frac{d-2}{d-3}T_+S_+-\frac{(d-2)\Omega_{d-2}}{16\pi}r_+^{d-3}$,
which upon using Eqs.~(\ref{gamma1}) and~(\ref{rplus}) yields 
$0= \frac{d-2}{d-3}T_+S_+-m$,
i.e., 
\begin{equation}
m= \frac{d-2}{d-3}T_+S_+,
\end{equation}
the Smarr relation for a black hole in $d$-dimensions, see
also \cite{smarrd}.
In four dimensions this is $m= 2T_+S_+$, the original Smarr formula
\cite{smarr}.

Smarr formula in $d$-dimensions has been provided before, but it is
remarkable that one can derive it from the shell mechanics and
thermodynamics in a non-trivial way.  The rest mass term $M$ that
surely appears in the Euler relation for the shell gives no
contribution, and it is the term $pA$ that contains the spacetime
energy and thus yields the mass $m$ term. This is in line with the black
hole mass formula derived in \cite{lz4} for $d=4$ using the membrane
paradigm approach, where the term $2p_+$ is indeed the usual black hole
surface gravity $\kappa$ divided by $4\pi$, and was  shown to be also
the horizon surface energy density $\sigma$ measured at infinity \cite{lz4}.
When $\sigma$ is multiplied by $A_+$ one obtains the total energy $m$.

\subsection{Other topics on black hole
entropy}

\subsubsection{Bekenstein 
entropy bound for the $d$-dimensional 
black hole}\label{dboundbh}

Let us now take the $d$-dimensional Bekenstein bound,  Eq.~(\ref{bekbound}),
in the black hole limit $R\to r_+$. For the
shell this bound is provided in Eq.~(\ref{bekbound1}) if we choose
$E=M$ and 
in Eq.~(\ref{bekbound2}) if we choose
$E=m$.

For $E=M$ the bound Eq.~(\ref{bekbound1}) is then 
\begin{equation}
S_E < 2S_+\,,
\end{equation}
where $S_+=\frac14\, A_+$ is the black hole entropy
given in Eq.~(\ref{SBH}).

For $E=m$ the bound Eq.~(\ref{bekbound2}) is 
\begin{equation}\label{bekbound2bh}
S_E <  S_+\,.
\end{equation}

Which case shall we choose, $E=M$ or $E=m$?
If we stick to $S_E < S_+$, i.e., to the statement
that the maximum
entropy for an area $A$ is when there is a
black hole in this area then
we should have 
Eq.~(\ref{bekbound2bh}) and so we should choose $E=m$,
see also \cite{Martinez}.
But all this this
relies on our previous choice of $\alpha$ and so the
argument is only of heuristic value.

\subsubsection{Holographic entropy bound for the $d$-dimensional 
black hole}\label{holdboundbh}

When the holographic entropy bound of Eq.~(\ref{Smax})
is applied to the shell we get 
Eq.~(\ref{Bbound1}). In the particular case that the shell
is a black hole then 
$a=0$ and $\eta = 1$ and we get
that the bound is satisfied for any $R\geqslant r_+$: This means
that in this case all the
shells, including the  black hole
limit, satisfy the bound. That the black hole
satisfies the bound is expected, since black holes
pose the highest entropy outcome from gravitational collapse.

\subsubsection{Entropy of the black hole for large $d$}

The entropy of a self-gravitating
shell for large $d$ is given in
Eq.~(\ref{sdlarge}). For the black hole case one puts
$a=0$ and $\eta=1$ to obtain
\begin{equation}\label{sdlargebh}
S_+ = \frac{1}{d^{d/2}}\, r_+^d\,.
\end{equation}
Note that $S_+$ is still 
$S_+=\frac14 A_+$
but in the large $d$ limit one has  $r_+\to 0$, $A_+\to 0$,
so the entropy of the black holes vanishes in the large $d$ limit.

\section{Conclusions}
\label{conc1}

The first law of thermodynamics on a $d$-dimensional self-gravitating
spherical thin shell is used in its entropy representation, where the
entropy is a function of the shell's rest mass and the shell's area
$A$, or since $A=4\pi R^2$, the shell's radius, $S=S(M,R)$.  The
pressure equation of state $p=p(M,R)$ is fixed by the spacetime
junction conditions, and the inverse temperature equation of state
$\beta=\beta(M,R)$ must have the form $\beta = k(r_+,R)\,b(r_+)$, with
$b(r_+)$ arbitrary, in order to satisfy the integrability condition
for the entropy, where $r_+=r_+(M,R)$. Integrating the first law, we
find that the shell's entropy is given as a function of the
gravitational radius $r_+(M,R)$ alone.

With the inverse temperature equation of state now controlled
completely by $b(r_+)$ we specify a power law equation for $b(r_+)$
with its exponent governed by a parameter $a$, and such that when
$a=0$ the inverse temperature $b(r_+)$ has the Hawking form.  The
thermodynamic stability conditions can be worked out generically, and
in particular, for $a=0$ it is found that the shell is stable when its
radius is in-between its own gravitational radius and the photonic
radius, i.e., the radius of circular photon orbits, reproducing
unexpectedly York's thermodynamic stability criterion for a $d=4$
black hole in a heat reservoir canonical ensemble.  Since the two
systems are different, this is an unexpected result, and hints that
what is important for thermodynamic stability is the place of the
shell alone, being it a heat reservoir massless shell or a massive
shell. An Euler formula for the matter is derived.

When put at its own gravitational radius the shell spacetime turns
into a black hole spacetime.  In this limit it is mandatory that the
self-gravitating shell is at the Hawking temperature which in turn
renders through the formalism developed the Bekenstein-Hawking entropy
in $d$ dimensions.  The black hole is marginally stable as the heat
capacity is zero. In this case the physical situation is the same as
in the $d=4$ York's case, York's heat reservoir shell and the massive
shell at the gravitational radius are the same thing, and so York's
criterion for marginal stability is precisely reproduced. The Smarr
formula for black holes pops out naturally and surprisingly.

\section*{Acknowledgments}
R.A. acknowledges support from the Doctoral Programme in the Physics 
and Mathematics of Information (DP-PMI) and the Funda\c c\~ao para a
Ci\^encia e Tecnologia (FCT Portugal) through Grant
No.~PD/BD/135011/2017.
J.P.S.L. acknowledges FCT for financial support through
Project~No.~UID/FIS/00099/2013, Grant No.~SFRH/BSAB/128455/2017, and
Coordena\c c\~ao de Aperfei\c coamento do Pessoal de N\'\i vel
Superior (CAPES), Brazil, for support within the Programa CSF-PVE,
Grant No.~88887.068694/2014-00.  G.Q. acknowledges the support of the
Funda\c c\~ao para a Ci\^encia e Tecnologia (FCT Portugal) through
Grant No.~SFRH/BD/92583/2013.

\end{document}